\documentclass[]{aastex631}

\usepackage[utf8]{inputenc}
\usepackage{amsmath}
\usepackage{graphicx}
\usepackage{subfigure}
\usepackage{amssymb}
\usepackage{amsthm}
\usepackage{xcolor}
\usepackage[figuresright]{rotating}
\usepackage{txfonts}
\usepackage{CJKutf8}
\usepackage{threeparttable}

\newcommand       \degree       {^\circ}

\shorttitle{The frequency-independent radio luminosity -- period relation of RS CVn stars}
\shortauthors{Huang et al.}

\graphicspath{{./}{figures/}}

\begin{document}
\begin{CJK}{UTF8}{gbsn}

\title{The frequency-independent radio luminosity -- orbital/rotational period relation of RS CVn stars}

\correspondingauthor{Qichen Huang,Biwei Jiang}
\email{qchuang@mail.bnu.edu.cn, bjiang@bnu.edu.cn}

\author[0000-0002-4046-2344]{Qichen Huang(黄启宸)}
\affiliation{Institute for Frontiers in Astronomy and Astrophysics, Beijing Normal University,  Beijing 102206, China}
\affiliation{School of Physics and Astronomy, Beijing Normal University, Beijing 100875, Peopleʼs Republic of China}
\affiliation{\textit{Current address:}School of Physics, The University of Sydney, Camperdown NSW 2050, Commonwealth of Australia}
\email{Current: qhua0119@uni.sydney.edu.au}

\author[0000-0003-3168-2617]{Biwei Jiang(姜碧沩)}
\affiliation{Institute for Frontiers in Astronomy and Astrophysics, Beijing Normal University,  Beijing 102206, China}
\affiliation{School of Physics and Astronomy, Beijing Normal University, Beijing 100875, Peopleʼs Republic of China}


\begin{abstract}

Radio emissions from RS CVn objects exhibit distinct characteristics at low and high frequencies, widely attributed to differing radiation mechanisms. The disparate processes of high-frequency gyrosynchrotron and low-frequency electron cyclotron maser emissions have traditionally suggested an absence of correlation in their radio luminosities. Our study presents a frequency-independent linear correlation between radio luminosity ($L_R$) and orbital/rotational periods ($P$) in RS CVn binaries. Analyzing the Sydney Radio Star Catalogue (SRSC) data, we derived orbital periods for 42 of 60 RS CVn sources using TESS light curves, revealing a strong positive correlation (PCC = 0.698, $P$ = 3.95e-7) between $\log_{10}L_R$ and $\log_{10}P$. This correlation remains across frequencies from 144-3000 MHz, showing uniform luminosity behavior. By combining light curve analysis with stellar mass-radius-luminosity relationships, we calculated parameters like binary mass, primary/secondary mass, Rossby number, and binary separation for eight RS CVn systems. The results show a notable correlation between radio luminosity and binary mass, primary mass, and separation (PCC = 0.663, 0.663, 0.719), with separation showing the strongest correlation. This suggests the radio emission may largely originate from the binary components' interaction, challenging existing models of RS CVn radio emission mechanisms and offering insights into the individual versus collective origins of these emissions.

\end{abstract}




\section{Introduction} \label{sec:intro}

RS CVn, fully designated as RS Canum Venaticorum variable, represents a class of binary stars first categorized by \citet{Hall1976}. These close binary systems exhibit intense chromospheric activity, with light curve variations caused by large-scale cool starspots. The prototype RS Canum Venaticorum is an eclipsing binary system. \citet{Strassmeier1990} classified its spectral type as F6IV+G8IV, indicating a double subgiant system, while \citet{Rodono2001} proposed an alternative classification of F5V+K2IV, suggesting a dwarf and subgiant pairing. Currently identified RS CVn systems encompass diverse stellar combinations ranging from double dwarfs to double giants. As noted by \citet{toet2021}, some systems consist of double late-type dwarfs (FGK spectral types), while \citet{Xiang2020} documented configurations featuring subgiant-main sequence pairs or double subgiants.

Typical characteristics of RS CVn objects include extensive starspot coverage (up to $30\%$ of the stellar surface) causing significant periodic optical luminosity variations. Differences in spot locations and formations yield distinct light curves across observation periods \citep{Eaton1979,Kang1989,Berdyugina2005,senav2018}. Component stars are tidally locked with short rotation and orbital periods ranging from under 1 day to 30 days \citep{Song2013,Reiners2014,toet2021}. These systems also display intense chromospheric and magnetic activity, enhancing luminosity across the electromagnetic spectrum from X-ray to radio wavelengths \citep{Garcia2003}.

Astronomers have conducted extensive radio observations of RS CVn systems due to their active nature. Radio luminosity varies significantly between systems, spanning $10^{14}$ to $10^{19}\ \rm {erg\ s^{-1}\ Hz^{-1}}$ in centimeter-wave observations \citep{Morris1988,Drake1989,Drake1992}. At GHz frequencies, emission typically exhibits long duration and low polarization, interpreted as gyrosynchrotron radiation. Rare instances of extended (hour-long) moderately polarized ($\sim 40\%$) emission that resembles planetary auroral processes are observed. At MHz frequencies, \citet{toet2021} confirmed through LOFAR observations at 144 MHz that the emission mechanism is coherent electron cyclotron maser emission, characterized by high polarization and brightness temperature.

Two fundamental questions regarding RS CVn radio emission remain unresolved. First is the origin of emission — whether from individual stellar components or magnetic structures connecting the binary system. \citet{beasley2000} observed quiescent emission linearly distributed over scales akin to binary separation, indicating large-scale magnetic structures. Likewise, \citet{Peterson2010} found that Algol's radio emission morphology aligns with the binary axis direction. In contrast, \citet{Bietenholz2012} concluded that radio emission from the RS CVn system IM Peg is confined to the primary star's disk-sized region, linking it exclusively to the primary star.
Second is the puzzling of the radio emission conform to the Güdel–Benz Relationship (GBR) across both high (GHz) and low (MHz) frequencies. The GBR is an empirical linear correlation between X-ray luminosity and radio luminosity observed at 5 GHz, featuring low-polarization incoherent gyrosynchrotron emission. This relationship is commonly attributed to radio emission and X-rays sharing a common energy source \citep{Benz1993,Benz2010,Gudel1993}. The 144MHz electron cyclotron maser radiation, due to the distinct nature of its emission mechanism, is not expected to conform to this relationship.
However, \citet{Vedantham2022} discovered that highly circularly polarized 144 MHz emission from RS CVn objects also follows the Güdel–Benz Relationship, challenging existing models of stellar coronal radio emission mechanisms.

Consequently, RS CVn systems present further exploratory opportunities in the radio band, not merely due to the interplay between radiation origin and mechanisms, but because all these mechanisms are linked to stellar magnetic fields. As interacting binaries, the high magnetic field strength of RS CVn largely stems from the interaction between the binary stars, meaning that the intrinsic properties of the stars and stellar parameters such as orbital separation and period significantly impact the magnetic fields. This implies an expected correlation between the stellar parameters of RS CVn objects and the radio emission intensity affected by magnetic fields. Therefore, we aim to advance the study of RS CVn characteristics by jointly analyzing radio data and stellar parameters.

Recent resources enable deeper investigation. The Sydney Radio Star Catalogue (SRSC) \citep{Driessen2024} contains 839 unique stars with 3,405 radio detections from MHz to GHz frequencies, more than doubling known radio stars. Stellar sources were identified through proper-motion matching and circular polarization searches \citep{pritchard2021,Driessen2024}, providing an unprecedented sample for systematic study. Additionally, the TESS light-curve data \citep{stassun2018tess} offers high-temporal-resolution monitoring (30-minute cadence, 2-minutes for preselected sources), particularly valuable for deriving parameters in short-period RS CVn systems.

In this work, we report a frequency-independent linear correlation between radio luminosity and orbital/rotational periods in RS CVn systems. This relationship necessitates reassessment of emission models predicting distinct GHz and MHz behaviors. We further analyze relationships between radio luminosity and stellar parameters—including individual masses, Rossby numbers, total system mass, and orbital semi-major axes—to investigate emission origins.

The paper is organized as follows: Section \ref{sec:Data} details sample selection and calculations of radio luminosity and periods. Section \ref{sec:Result} presents the period-luminosity relationship with statistical validation. Section \ref{sec:Discussion} examines correlations with stellar and binary parameters. Section \ref{sec:Summary} provides conclusions and future prospects.

\section{Data} \label{sec:Data}

\subsection{Sample} \label{sample}

Our data is sourced from the Sydney Radio Star Catalogue (SRSC) published by \citet{Driessen2024}, which contains 839 stars with 3,405 radio observation records spanning frequencies from 144 MHz (LOFAR) to 3000 MHz (VLA). The majority of these observations come from the Rapid ASKAP Continuum Survey (RACS) conducted by the Australian SKA Pathfinder (ASKAP), covering three bands: RACS-low (887.5 MHz), RACS-mid (1367.5 MHz), and RACS-high (1655.5 MHz). Additional limited data come from ASKAP observations at 855.5 MHz, 943.5 MHz, and 1013.5 MHz, along with the MeerKAT observations at 1284 MHz.

Since SRSC is cross-matched with SIMBAD, we directly filtered RS CVn objects using SIMBAD's object type, yielding 60 sources. While many unidentified RS CVn stars likely exist among eclipsing binaries, we prioritized sample purity by retaining only sources explicitly classified as RS CVn in SIMBAD's main-type designation.

\begin{figure}
    \centering
    \includegraphics[scale=0.55]{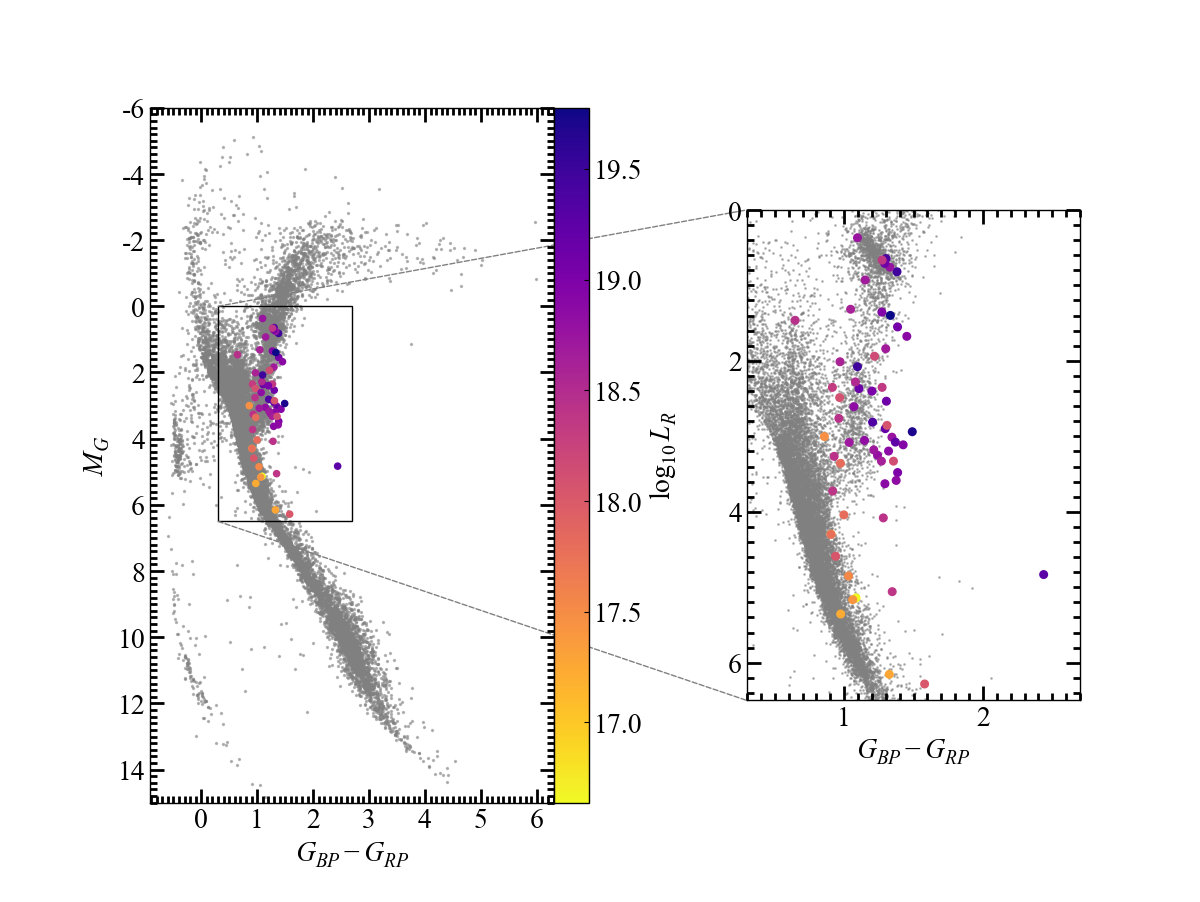}
    \caption{60 RS CVn stars placed in a Gaia third data release (DR3) color–magnitude diagram (CMD). Gray points are 2-minute cadence observation targets of TESS sectors 1 and 2 which also contained in Gaia DR3. All distance data were provided by \citet{Bailer-Jones2021}. Colors indicated the maximum radio luminosity of RS CVn stars.}
    \label{fig:CMD}
\end{figure}

For these 60 RS CVn systems, Gaia DR3 provides photometric data in the $G$, $G_{\rm BP}$, and $G_{\rm RP}$ bands. Due to limited spatial resolution, these measurements represent combined photometry for both binary components in the RS CVn system. Using precise distances from \citet{Bailer-Jones2021} based on Gaia EDR3, the $G$-band photometry was converted to absolute magnitude $M_{\rm G}$. Following \citet{Driessen2024}'s approach,  a color-magnitude diagram (CMD) was constructed and shown in Figure \ref{fig:CMD} by combining the 60 RS CVn systems with background sources from TESS Sectors 1 and 2 of the 2-minute cadence targets. In the figure, the 60 RS CVn sources are extinction corrected using the extinction values calculated by PYSSED\footnote{The Python Stellar Spectral Energy Distribution. https://github.com/iain-mcdonald/PySSED}. Only one source exhibited significant extinction with $E(G_{\rm BP}-G_{\rm RP}) = 0.16$, while 53 sources had negligible extinction with $E(G_{\rm BP}-G_{\rm RP}) < 0.01$. Background sources were included without extinction correction.

\subsection{Radio luminosity}

The radio luminosities for all RS CVn sources were derived from the observational data in the SRSC catalog. For radio point sources, the integrated flux density should equal to the peak flux density, and the sources in our data that contain both measurements generally satisfy this condition. When calculating radio luminosities, we prioritized using the integrated flux density. For three sources lacking integrated flux density data, their peak flux density were used for calculations. Given that most sources in SRSC have multiple observations, we adopted the maximum flux density for each source following \citet{Driessen2024}'s methodology. The corresponding maximum radio luminosity was then computed using distances from \citet{Bailer-Jones2021} based on Gaia EDR3.

Flux densities exhibit substantial variation across sources due to distance differences, ranging from 0.23 mJy to 94.9 mJy. Approximately 50\% of values fall between 1–5 mJy, and 80\% between 0.5–10 mJy. Source distances span 27 pc to 688 pc, with roughly 50\% within 120 pc and 90\% within 300 pc. These predominantly short distances align with the previously noted low extinction. All data are presented in Table \ref{tab:RS Cvn radio data}.

\subsection{Period}

To obtain the rotation periods of the 60 RS CVn sources, the Transiting Exoplanet Survey Satellite (TESS) photometric data were used. The TESS Input Catalog (TIC) currently contains over 1 billion objects \citep{stassun2018tess}. The light-curve data are extracted from MIT’s Quick-Look Pipeline (QLP)\citep{Huang2020a,Huang2020b,Kunimoto2021,Kunimoto2022}, which covers the full two-year TESS Primary Mission. This dataset includes approximately 14.77 million and 9.6 million individual light-curve segments in the southern and northern ecliptic hemispheres, respectively.

Period determination from TESS light curves relies on pronounced brightness variations induced by large starspot regions during stellar rotation. Although rapid spot evolution causes variations between cycles, these systems maintain periodic characteristics corresponding to their rotation period. Given the tidal locking in RS CVn systems, rotation periods are nearly identical to orbital periods, enabling simultaneous determination of both.

Of the 60 RS CVn sources, the TESS light curve data were obtained for 57  after cross-matching with the TESS Input Catalog v8.2 \citep{Paegert2021} and QLP database using Gaia coordinates with a 1$\arcsec$ radius. To identify periodic variations, the widely used Lightkurve package and LombScargle in Astropy are taken to generate the periodograms and light curves. Since the periodic signals were  prominent in most sources, only a constraint of signal-to-noise ratio greater than 10 was imposed. Given that RS CVn stars show significant variations in different periods, the one with the best observational quality was typically selected for period determination  when multiple observational datasets from different epochs were available.

Periods were successfully determined for 42 sources, ranging from 0.31 to 18.5 days. Approximately 60\% fall between 1-5 days, and 80\% between 1-10 days, consistent with typical RS CVn periods. We verified each period through phase folding and calculated standard deviations from the full width at half maximum (FWHM) of the periodogram. Relative errors were below 10\% for $\sim$90\% of sources, with a maximum error of 18.75\%. Fifteen sources were excluded due to insufficient periodic signals. All period data and associated errors are listed in Table \ref{tab:RS Cvn radio data}.

\section{Result} \label{sec:Result}

\begin{figure}
    \centering
    \subfigure[]{\includegraphics[width=0.62\textwidth]{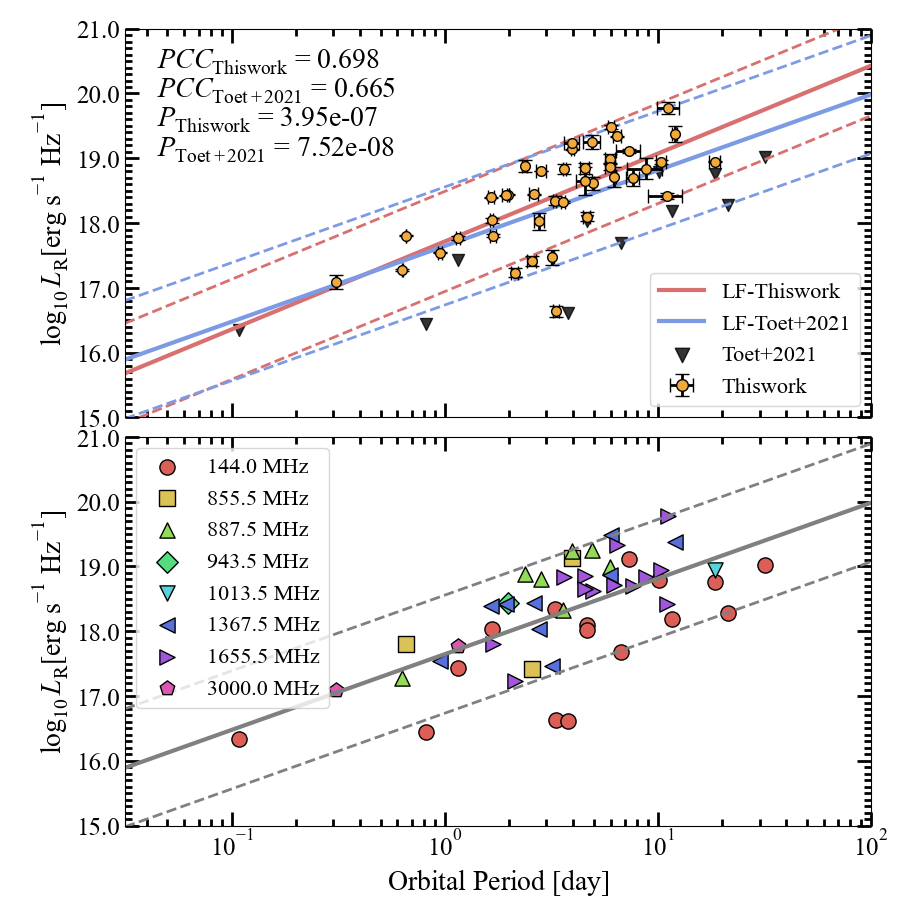}}
    \\
    \subfigure[]{\includegraphics[width=0.65\textwidth]{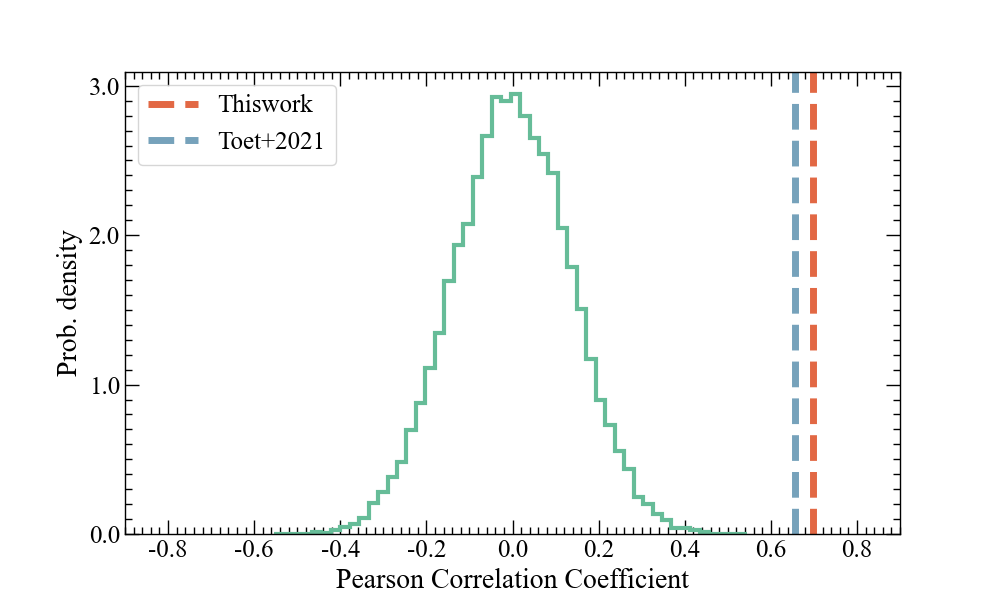}}
    \caption{(a)Upper panel: Maximum radio luminosity - Rotation period correlation, where circles represent sources from SRSC and black triangles represent sources from \citet{toet2021}. The red solid line (LF-Thiswork) represents the linear fit to 42 RS CVn stars in our work and the blue solid line (LF-Toet+2021) represents the linear fit to 54 RS CVns which included 12 sources from \citet{toet2021}. The dashed lines of each color indicating the $1.5\sigma$ confidence intervals of the fit. The Pearson correlation coefficient of both 42/54 sources is listed in the upper left. The value of $P$ indicates the probability of observing a PCC value equal to the calculated result, under the null hypothesis that there is truly no linear correlation between the two variables in the population. $\ $
    Lower panel: Maximum radio luminosity - Rotation period correlation for all 54 sources, with observations from 144 MHz to 3000 MHz represented by different colors and shapes. The meaning of solid and dashed lines are the same as those in the upper panel.
    (b) The distribution of Pearson Correlation Coefficients (PCC) between orbital period and radio luminosity derived from 1000 sets of MCMC simulations, with the red dashed line indicating the PCC value of 0.698 calculated from our 42 sources.}
    \label{fig:PCC&MCMC}
\end{figure}

Figure \ref{fig:PCC&MCMC} (a) displays the linear relationship between maximum radio luminosity and rotation/orbital period for 42 RS CVn sources (red line). The error bars on the x-axis indicate period standard deviations calculated from the full width at half-maximum (FWHM) of the periodogram, while the y-axis error bars represent radio luminosity uncertainties propagated from flux density and distance measurements. A significant linear correlation is evident between the logarithmic period ($\log{P}$) and the logarithmic radio luminosity ($\log_{10}{L_{\rm R}}$). Validation through Pearson correlation coefficient (PCC) analysis yields PCC = 0.698 with $P = 3.95\times10^{-7}$, indicating the probability of no correlation between these parameters is only about $10^{-7}$.

\subsection{Incorporating Low-Frequency Data}

To augment the low-frequency (144 MHz) dataset, the parameters for 14 RS CVn radio sources identified by \citet{toet2021} were incorporated. Three sources overlapped with our original sample of 42 objects: DM Uma and EV Dra (observed at the same frequency), and BH CVn (observed at different frequencies). For DM Uma and EV Dra, the SRSC data are retained due to their higher flux densities. For BH CVn,  both SRSC and \citet{toet2021} data are reserved because their different observational frequencies may suggest distinct emission mechanisms. This resulted in 12 new sources being added to our analysis.

To verify the consistency with the previously identified period–radio luminosity relationship, we calculated the Pearson correlation coefficient for the combined dataset (42 original plus 12 new sources), shown in the upper panel of Figure \ref{fig:PCC&MCMC} (a). The resulting PCC remained strong at 0.665 with $P = 3\times10^{-8}$. Although slightly lower than our initial coefficient (0.698), this value maintains statistical significance and aligns with our core finding.

\subsection{Frequency Independence}\label{subsec:Frequency}

To investigate whether observational frequencies affect the overall radio luminosity distribution, the observing frequencies are annotated for all data points in the lower panel of Figure \ref{fig:PCC&MCMC} (a), including the 12 RS CVn sources from \citet{toet2021}. The frequencies exhibit no significant regional clustering in the diagram. Only the 144 MHz data displays a slight systematic offset toward lower values compared to other frequencies at similar periods.

As noted previously, RS CVn systems have different radio emission mechanisms at high and low frequencies. 
At GHz frequencies, gyrosynchrotron radiation from mildly relativistic electrons, injected into and trapped within closed magnetic fields, is the predominant incoherent radio emission mechanism for RS CVn \citep{Benz2010}. For instance, research by \citet{Kuijpers1985} suggested that radiation from the $\sigma^2\ CrB$ system could be explained as synchrotron emission from electrons with Lorentz factors of 3.5-3.1 in magnetic fields of strength 27-35 Gauss. Similarly, \citet{Gudel2002} proposed gyrosynchrotron emission as the radiation mechanism for the HR 1099 system based on its radio spectrum from 1.5 GHz to 10 GHz. Gyrosynchrotron radiation typically exhibits low levels of circular polarization, with a radio spectrum displaying a negative slope, and its intensity is related to the magnetic field strength and the distribution of electron energies. 

At MHz frequencies, \citet{toet2021} confirmed through LOFAR observations at 144 MHz that the radio emission mechanism for RS CVn systems at this frequency is coherent electron cyclotron maser emission (ECM). This process is driven by the horseshoe or shell-type plasma instability within stellar coronae (or planetary magnetic fields), which are highly sensitive to gradients in electron momentum distribution, in contrast to total electron energy, for gyrosynchrotron emission.
\citet{White1995} also noted the contrasting polarization at high and low frequencies in RS CVn systems, attributing it to different radiation mechanisms: incoherent gyrosynchrotron emission at high frequencies and a form of coherent plasma emission at low frequencies. ECM typically exhibits strong circular polarization and boasts brightness temperatures far exceeding that of thermal radiation. 
For RS CVn systems with a magnetic field strength around 100 Gauss, this mechanism results in a pronounced cut-off effect when the observation frequency exceeds the electron cyclotron frequency (approximately 300 MHz). Consequently, the radio emission mechanism observed at GHz frequencies must necessarily differ from that at 144 MHz.

Notably, our sample spans frequencies from 144 MHz to 3000 MHz, theoretically encompassing both emission mechanisms. However, these distinct mechanisms show no differential distribution in the period–radio luminosity relationship.

\subsection{Verification through Markov Chain Monte Carlo (MCMC)}

Adopting the methodology of \citet{Vedantham2022}, we employed Markov Chain Monte Carlo (MCMC) techniques to generate 1000 simulated datasets of radio luminosity and orbital period pairs. This analysis tests whether the observed correlation could arise from chance occurrences or statistical bias under the null hypothesis that radio luminosity and orbital period are independent random variables.

Given the diverse telescope origins of SRSC data and the uncertain space density of RS CVn systems, we adopted a 1mJy observational threshold. Stellar distances were randomly distributed (non-Gaussian) between a minimum of 27pc and a maximum of 385pc (only one source among the 42 sources had a distance greater than 385pc at 688pc, which significantly deviates from the majority).

The radio luminosity was generated following a normal distribution with a mean of $10^{18.5}\ \rm erg\ s^{-1}\ Hz^{-1}$ and standard deviation of $10^{0.5}\ \rm erg\ s^{-1}\ Hz^{-1}$. The orbital period followed a normal distribution with a mean of $10^{0.5}\ \rm days$ and standard deviation of $10^{0.2}\ \rm days$. Each iteration generated 55 sources (slightly more than our sample size, considering the average number of simulated observational stars is close to our sample), retaining only those sources whose calculated radio flux density exceeded the detection threshold based on distance. The Pearson correlation coefficient between radio luminosity and orbital period was calculated for each dataset.

Figure \ref{fig:PCC&MCMC} (c) displays the distribution of Pearson correlation coefficients for the 1000 MCMC simulations. Only 0.003$\%$ of the simulated datasets achieved a PCC$\ge$0.5, while our observed result shows a much stronger correlation with $\rm PCC=0.698$. This demonstrates that our results cannot be attributed to coincidence or observational bias.


\section{Discussion} \label{sec:Discussion}

To investigate potential factors affecting the radio luminosity-period relationship and probe the origin of radio emission in RS CVn systems, we performed a comprehensive analysis of stellar and orbital parameters, examining how individual component properties and binary separation may influence the overall radio luminosity of the systems.

\subsection{Stellar parameter calculation} \label{subsec:para calculation}

To identify the dominant factors influencing radio luminosity, several stellar parameters were calculated. These include individual stellar parameters — the primary and secondary masses ($M_{\rm pri}$, $M_{\rm sec}$) and Rossby number ($R_{\rm o}$), and the binary parameters: total system mass ($M_{\rm tot}$) and orbital semi-major axis ($a$).

However, the close binary nature of RS CVn systems prevents most observational data from resolving emission from individual components. This complicates parameter analysis through traditional methods like stellar color indices. We therefore employed a three-step approach:
\begin{enumerate}
\item Performing spectral energy distribution (SED) fitting using PySSED to determine the system's total luminosity;
\item Deriving component temperature ratios from the TESS light curve data;
\item Calculating individual component effective temperatures using the total luminosity and temperature ratios obtained above.
\end{enumerate}

\subsubsection{SED fitting and  effective temperature ratio calculation}

The Python Stellar Spectral Energy Distribution (PYSSED) package combines photometry from disparate catalogs to fit stellar parameters including luminosity, temperature, extinction, and other characteristics \citep{McDonald2024}. Since photometric data generally cannot resolve individual components in RS CVn systems, PySSED treats the binary as a single source during fitting. Consequently, the derived parameters like temperature, stellar radius, and surface gravity are unreliable. However, the luminosity, being a combined contribution from both stars, remains robust against this  limitation. Thus, the total system luminosity is obtained for 41 of 42 sources with one excluded due to insufficient photometric data using PySSED. The luminosities span from 0.31$\rm L_{\rm \odot}$ to 66.72$\rm L_{\rm \odot}$, with 31 sources ($\sim75\%$)  below 10$\rm L_{\rm \odot}$ and 12 sources ($\sim25\%$) below 3$\rm L_{\rm \odot}$. The former threshold (10$\rm L_{\rm \odot}$) is approximately half of the minimum luminosity for giant stars, while the latter (3$\rm L_{\rm \odot}$) is about twice the total luminosity of an F8 dwarf star.

Binary light curves exhibit transit phenomena where stellar components overlap along the line of sight, reducing overall brightness. Despite identical occulted areas during mutual eclipses, the luminosity decrease magnitude varies due to temperature differences between components. Following the Stefan-Boltzmann law, the ratio of eclipse depths is proportional to the effective temperature ratio ($T_{\rm eff, sec}/T_{\rm eff, pri}$). We therefore determined temperature ratios by measuring relative luminosity decrease amplitudes.

Extensive starspot coverage in RS CVn systems complicates transit measurements, as eclipse signals superimpose on rotational variability. To address this, we first removed rotational periodic signals using Lightkurve, then performed numerical measurements on phase-folded residuals.

\begin{figure}
    \centering
    \subfigure[]{\includegraphics[scale=0.45]{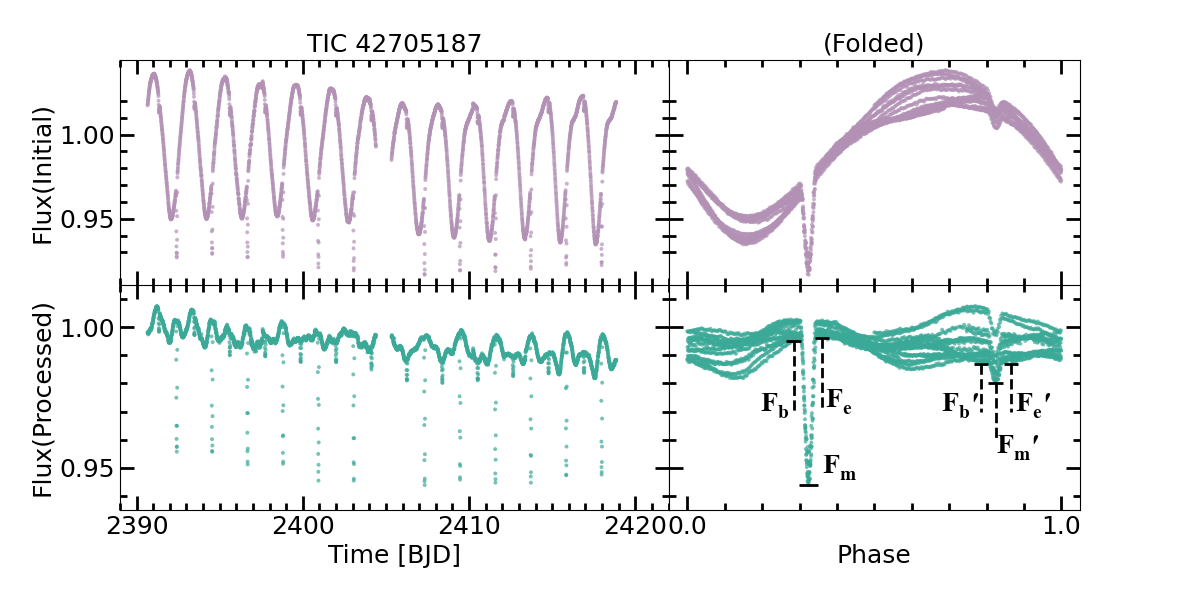}}
    \subfigure[]{\includegraphics[scale=0.45]{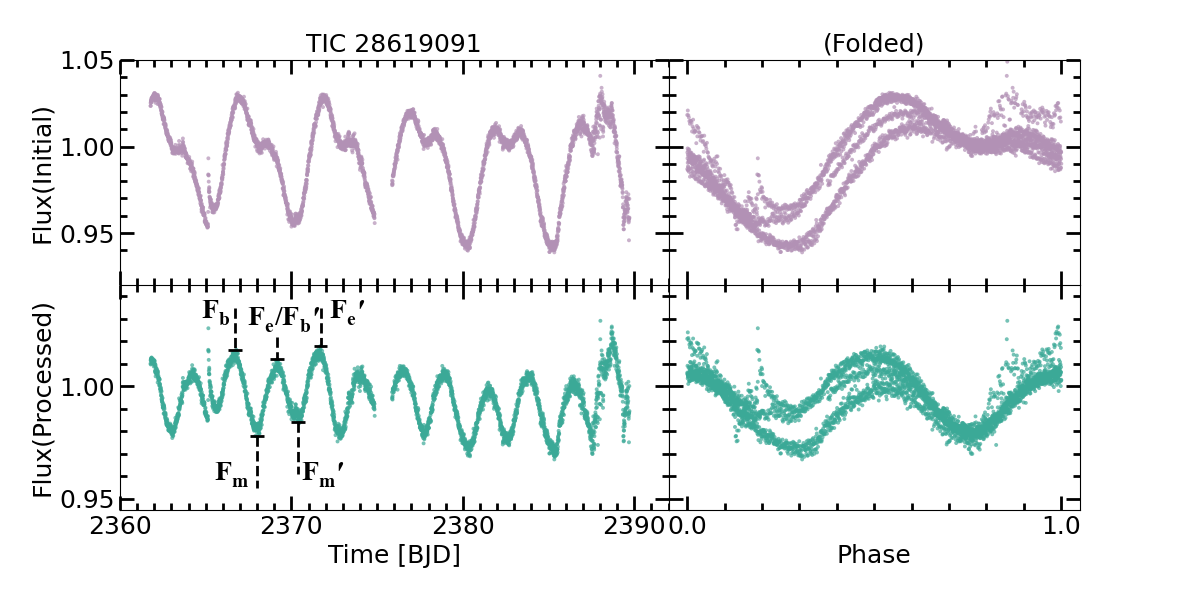}}
    \caption{Light curves for (a) TIC 42705187 (2.13-day period) and (b) TIC 28619091 (4.9-day period). For each source, upper panels show original light curves; lower panels display results after periodic signal removal (strongest signal removed iteratively 4 times for (a), once for (b)). Left panels present time-series data; right panels show phase-folded results. Eclipse depth measurements used the lower-right panel for (a) and lower-left panel for (b).}
    \label{fig:lightcure-all}
\end{figure}

Figure \ref{fig:lightcure-all} (a) shows the light curves of TIC 42705187 used for measurements. The top two panels display the data before periodic signal removal, while the bottom two panels show the data after iteratively removing the strongest periodic signal 4 times. The left panels present the original light curves, and the right panels show the results after phase-folding with a 2.13 day period.

Eclipse depths were calculated by averaging flux values at the start ($F_b$, $F_b'$) and end ($F_e$, $F_e'$) of each transit, then subtracting the corresponding minima ($F_m$, $F_m'$). The depth ratio is given by:

\begin{equation}
    \begin{aligned}
        E_1 : E_2 = \left[ \frac{F_b + F_e}{2} - F_m \right] : \left[ \frac{F_b' + F_e'}{2} - F_m' \right]
    \end{aligned}
    \label{eq:steven-bolzemann}
\end{equation}

Since RS CVn light curves cannot achieve a completely flat state even after multiple periodic signal removals, this computational method helps minimize the effects of overall brightness variations.

Some sources (e.g., Figure \ref{fig:lightcure-all}b) exhibit eclipse signals masked by phase-modulated variability. After removing the dominant periodic signal, these display two transits per period — characteristic of contact binaries \citep{Debski2022, Guo2022}. As described by \citet{Eggleton2012}, contact binaries feature extremely close components ($\sim$0.01 AU separation) with short periods (1–10 days), forming a peanut-shaped shared envelope. RS CVn systems frequently exhibit this configuration.

Their light curves prove more complex than typical contact binaries due to spot-induced luminosity variations. Even after phase folding, identifying primary/secondary peaks remains challenging (Figure \ref{fig:lightcure-all}b). For such sources, we derived eclipse amplitudes by comparing both transits within one period of the processed light curve, minimizing spot-induced brightness variation effects.

We successfully calculated effective temperature ratios for 32 of 42 sources, ranging from 0.63 to 0.99. Twenty sources ($\sim50\%$) show ratios $>0.9$ (indicating similar component temperatures), while seven ($\sim20\%$) exhibit ratios $<0.8$ (suggesting significant temperature differences).

\subsubsection{Calculation of $T_{\rm eff}$, Mass and Radius for Dwarf-Dwarf Systems}

Given that RS CVn systems may contain dwarf, subgiant, or giant components, we classified systems with total luminosity ≤3 $\rm L_{\odot}$ as dwarf-dwarf systems for separate analysis. Among our 42 sources, 8 satisfy this luminosity criterion and have determined effective temperature ratios, enabling calculation of component $T_{\rm eff}$, mass, and radius.

We adopted the Mass-Luminosity Relation (MLR), Mass-Effective Temperature Relation (MTR), and Mass-Radius Relation (MRR) for dwarf stars (0.179-31 $\rm M_{\odot}$) from \citet{Eker2018}. Since RS CVn systems lack high-temperature stars, we restricted the relation for stars with $\le$ 1.4 $\rm T_\odot$. Combining MTR and MRR yields the Effective Temperature-Radius Relation (TRR). For stars $\le$ 1.5 $\rm M_{\odot}$, MTR must be derived jointly from MLR and MRR. Below 1.4 $\rm T_{\odot}$, we divided the TRR into three segments (excluding the 0.570-0.610 $\rm T_{\odot}$ range as inapplicable):


\begin{equation}
    \begin{aligned}
        R = 2.224\times T^2 -1.761\times T +0.622 \qquad [0.610-0.798\ \rm T_{\rm \odot}]
        \\
        R = 1.212\times T^2 -0.629\times T +0.410 \qquad [0.798-1.036\ \rm T_{\rm \odot}]
        \\
        R = 1.841\times T^2 -1.968\times T +1.126 \qquad [1.036-1.400\ \rm T_{\rm \odot}]
    \end{aligned}
    \label{eq:TRR}
\end{equation}

Applying the Stefan-Boltzmann law, we established:

\begin{equation}
    \begin{aligned}
        L_{\rm tot} = R_1^2 \times T_1^4 + R_2^2 \times T_2^4 \\
        R_{\rm T} \times T_1 = T_2
    \end{aligned}
    \label{eq:Luminosity-TR}
\end{equation}
where $L_{\rm tot}$ is the PySSED-derived total luminosity, $R_1$/$R_2$ and $T_1$/$T_2$ are component radii and effective temperatures, and $R_{\rm T}$ is the measured temperature ratio.

Combining Equations \ref{eq:TRR} and \ref{eq:Luminosity-TR}, an equation with $L_{\rm tot}$ on the left side and terms composed of $R_{\rm T}$ and $T_1$ on the right side could be derived. This allows us to calculate the effective temperatures of both stars using known values of $L_{\rm tot}$ and $R_{\rm T}$. Based on previous calculations and using Effective Temperature-Radius Relation (TRR) and Mass-Effective Temperature Relation (MTR), the effective temperatures, radii, and stellar masses for both components in all 8 RS CVn dwarf-dwarf systems were determined.

However, this sample size (n=8) presents limitations. Most RS CVn systems in our catalog contain subgiants—objects in brief evolutionary phases lacking established temperature-luminosity-mass relations. 

To address this, we incorporated parameters for 14 additional RS CVn systems from \citet{toet2021}. EV Dra overlapped between samples, enabling validation; Our period (1.66 days vs. 1.672 days), semi-major axis (0.033 AU vs. 0.029 AU), and primary mass (0.961 $\rm M_{\odot}$ vs. 0.98 $\rm M_{\odot}$) show excellent agreement. The secondary mass discrepancy (0.872 $\rm M_{\odot}$ vs. 0.82 $\rm M_{\odot}$) remains modest. This consistency validates our methodology, so we retained our EV Dra parameters for subsequent analyses. Parameters for the remaining 13 systems from \citet{toet2021} are listed in Table \ref{tab:Stellar parameter}.

\subsubsection{Calculation of the Rossby number}

Among individual stellar parameters, stellar activity stands out as a crucial aspect requiring examination, because for late-type stars, there exists a correlation between activity and radio emission capability \citep{yiu2024}. \citet{wright2011} studied the relationship between the activity and rotation rates of 824 solar-type and late-type stars. They found that the activity quantified by X-ray-to-bolometric luminosity correlates with stellar rotation rates and can be parameterized by the Rossby number $R_{\rm o} \equiv P_{\rm rot}/\tau$. Here, $P_{\rm rot}$ denotes the rotation period and $\tau$ the convective turnover time, which is derivable from stellar mass. Following this approach, \citet{Huang2024} employed $R_{\rm o}$ to identify highly active dwarf stars as potential radio sources.

We computed $\tau$ using the mass-dependent relation from \citet{wright2011}:
\begin{equation}
    \begin{aligned}
        \log_{10}\tau = 1.16-1.49\log_{10}{(M/M_\odot)}-0.54\log_{10}^2{(M/M_\odot)}
    \end{aligned}
    \label{eq:steven-bolzemann}
\end{equation}

The Rossby numbers $R_{\rm o}$ were calculated for 35 stellar components: 8 RS CVn systems from our calculations and 10 systems from \citet{toet2021} (one lacking secondary mass data). The Rossby numbers range from 0.02 to 17.95, with 18 individual stars having $R_{\rm o} \le$ 0.13. In 10 of the 18 RS CVn systems, at least one component in each system exhibits $R_{\rm o} \le$ 0.13.

\subsubsection{Calculation of semi-major axis}

Among all derivable binary parameters, the orbital semi-major axis (binary separation) ranks as one of the most critical characteristics, second only to the orbital period. As a parameter independent of single-star properties, it provides an ideal metric for investigating correlations with the total radio luminosity of RS CVn systems. We computed the orbital semi-major axes for our sample of 8 RS CVn systems using Kepler's Third Law, which relates the orbital period $(P)$ and total binary mass $(M_{\rm tot})$ through:
$P^2 = {a^3}/{M_{\rm tot}}$.
This relation can be expressed logarithmically as $( 2\log_{10}{P} + \log_{10}{M_{\rm tot}} = 3\log_{10}{a} )$. The calculated results, listed in Table \ref{tab:Stellar parameter}, are fully consistent with the characteristic scale of \(\sim\)0.01 AU reported by \citet{toet2021}.

For the 13 sources in \citet{toet2021}, orbital semi-major axis values were directly available for 10 systems. Among these, eight were compiled from literature, while the remaining two were derived by \citet{toet2021} using Kepler's Third Law. Combining these with our sample yields a total of 18 RS CVn systems with known orbital semi-major axes, spanning 0.011–0.315 AU, of which 14 systems exhibit separations below 0.1 AU.

\subsection{The relationship between stellar parameters and radio luminosity}\label{subsec: para and radio}

\subsubsection{Individual stellar parameters}

\begin{figure}
    \centering
    \subfigure[]{\includegraphics[width=0.6\textwidth]{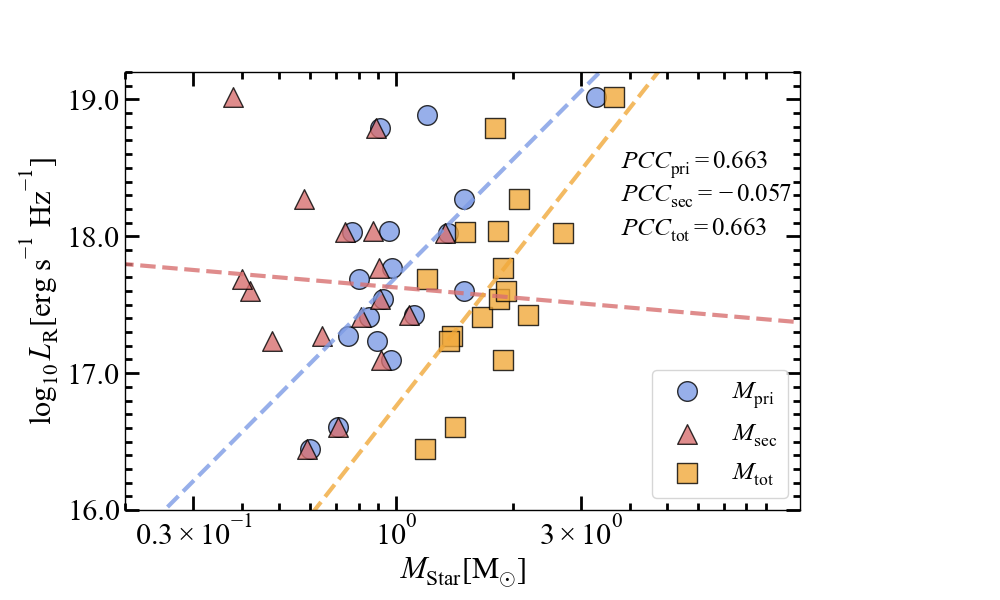}}
    \\
    \subfigure[]{\includegraphics[width=0.6\textwidth]{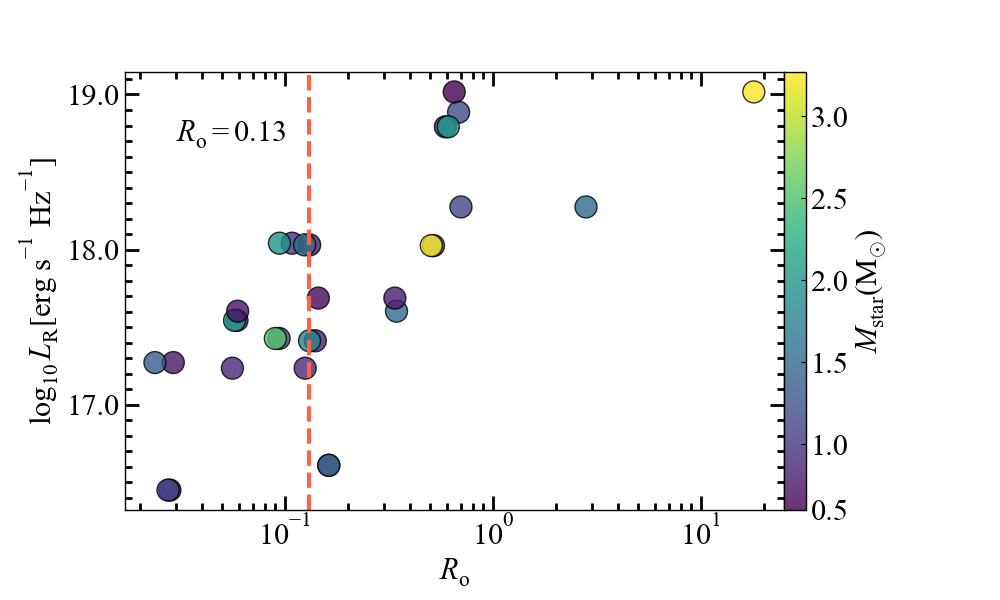}}
    \\
    \subfigure[]{\includegraphics[width=0.6\textwidth]{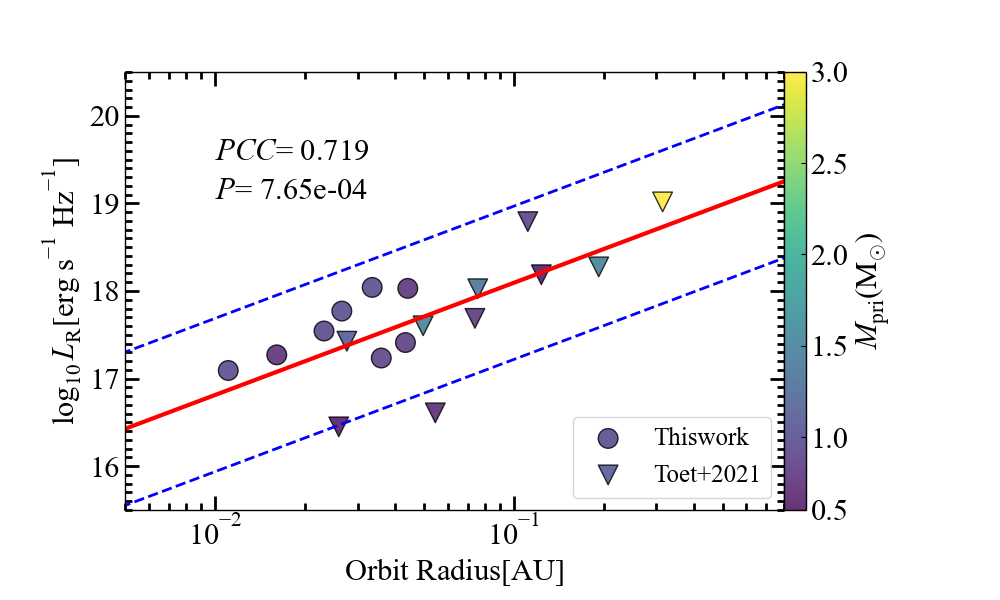}}
    \caption{
        (a) Relationship between stellar mass and maximum radio luminosity. Red triangles denote secondary stars, blue circles primary stars, and yellow squares total binary mass. Colored dashed lines show linear fits for each mass category.
        (b) Relationship between Rossby number (activity indicator) and maximum radio luminosity, with color coding indicating stellar mass. The dashed line at $R_{\rm o}=0.13$ marks the stellar activity boundary.
        (c) Relationship between binary semi-major axis and maximum radio luminosity. Circles represent SRSC sources, triangles denote sources from \citet{toet2021}, with colors indicating primary star mass.
    }
    \label{fig:Discussion Mass-Ro-Radius}
\end{figure}

We analyzed individual stellar parameters, focusing on stellar masses and Rossby numbers derived from our calculations. Figure \ref{fig:Discussion Mass-Ro-Radius} (a) displays relationships between radio luminosity and stellar masses (primary, secondary, and total binary mass) in RS CVn systems. Primary stars are denoted by blue circles, secondary stars by red triangles, and total binary mass by yellow squares. Blue, red, and yellow dashed lines represent linear fits for primary, secondary, and total masses respectively, with the x-axis in logarithmic scale.

Pearson correlation coefficients (PCC) between $\log_{10}(M_{\rm Star})$ and radio luminosity are shown in the upper right corner. Results indicate no correlation between secondary star mass and radio luminosity (PCC $\approx$ 0). In contrast, both primary star mass and total binary mass exhibit significant correlations with total radio luminosity, with nearly identical correlation strengths (PCC = 0.663).

This suggests that secondary stars in RS CVn systems contribute minimally to total radio emission intensity, with primary stars being the dominant source. However, the binary system as a whole may also influence RS CVn radio luminosity.


Figure \ref{fig:Discussion Mass-Ro-Radius} (b) presents the relationship between Rossby number ($R_{\rm o}$) and radio luminosity for 35 individual stars. The red dashed line indicates $R_{\rm o}=0.13$, the established boundary for stellar activity saturation. Based on \citet{fang2018,anthony2022}, stellar activity saturates when $R_{\rm o} \leq 0.13$, while for $R_{\rm o} > 0.13$, activity increases as $R_{\rm o}$ decreases. We thus investigated whether the more active star significantly determines radio luminosity.

Figure \ref{fig:Discussion Mass-Ro-Radius} (b) reveals a weak positive correlation between radio luminosity and $R_{\rm o}$, with higher radio luminosity corresponding to larger $R_{\rm o}$ values (lower activity). This implies that RS CVn radio luminosity cannot be primarily attributed to individual stellar component activity.

\subsubsection{Binary parameters}

For binary parameters, we focused on analyzing the orbital semi-major axis. As previously mentioned, this parameter was calculated using Kepler's Third Law based on the orbital period $P$ and total binary mass $M_{\rm tot}$, expressed as: $2\log_{10}{P} + \log_{10}{(M_{\rm tot})} = 3\log_{10}{a}$.
In this formulation, both logarithmic terms on the left side ($\log_{10}{P}$ and $\log_{10}{M_{\rm tot}}$) exhibit positive linear correlations with radio luminosity. Consequently, the logarithmic semi-major axis ($\log_{10}{a}$) naturally demonstrates a positive linear correlation with radio luminosity.

However, in our study, the orbital period $P$ and total binary mass $M_{\rm tot}$ are mutually independent parameters with no intrinsic correlation. Theoretically, each semi-major axis $a$ corresponds to infinitely many $(P, M_{\rm tot})$ combinations. We therefore conclude that at least two of the three parameters ($P$, $M_{\rm tot}$, $a$) exhibit genuine positive linear correlations with RS CVn radio luminosity, while the third parameter's apparent correlation may arise indirectly through its mathematical relationship with the other two.

Figure \ref{fig:Discussion Mass-Ro-Radius} (c) displays the relationship between orbital semi-major axis and radio luminosity for 18 RS CVn systems. A significant positive linear correlation (PCC = 0.719) emerges, stronger than those observed for either orbital period $P$ or total binary mass $M_{\rm tot}$ individually. This suggests that binary separation may be a more critical determinant of RS CVn radio luminosity. Furthermore, the color distribution in Figure \ref{fig:Discussion Mass-Ro-Radius} (c) indicates no strict correlation between semi-major axis and primary star mass, implying that the semi-major axis-radio luminosity correlation exists independently of the primary mass-radio luminosity relationship.

\vspace{1cm}

Based on these findings, we conclude that while the primary star significantly contributes to RS CVn radio luminosity, the combined effect of the binary system likely equals or exceeds this contribution, whereas the secondary star has negligible impact. This framework partially reconciles the differing observations of \citet{Peterson2010} and \citet{Bietenholz2012}: although the primary star dominates, extended structures associated with the binary system as a whole contribute to emission, completely independent of the secondary star.

\section{Summary}\label{sec:Summary}

Based on a total of 60 RS CVn systems provided in the Sydney Radio Star Catalogue (SRSC), this study derived rotation/orbital periods for 42 of these systems through analysis of TESS light curves, thereby establishing a frequency-independent positive correlation between maximum radio luminosity and rotation/orbital period. This relationship persists despite distinct emission mechanisms at different frequencies, i.e. electron cyclotron maser emission at MHz frequencies versus gyrosynchrotron emission at GHz frequencies.

Through SED fitting using PYSSED, TESS light curve analysis, and the application of stellar parameter formulas from \citet{Eker2018}, stellar parameters for 8 dwarf-dwarf systems among the 42 sources were calculated, including binary temperatures, masses, Rossby numbers, and orbital semi-major axes.

To expand the dataset, this study incorporated data from \citet{toet2021}, which included 14 RS CVn radio sources. It should be noted that due to partial data overlap, different numbers of sources from \citet{toet2021} were introduced as supplements in various sections of this study. 
In the orbital period-radio luminosity-frequency section, 3 of the 14 sources overlapped with the 42 sources in this study; 2 had identical observation frequencies to those in SRSC (thus excluded), while 1 had different observation frequencies (thus retained), ultimately introducing 12 additional sources. 
In the stellar mass, Rossby number, and orbital radius sections, 1 of the 14 sources overlapped with the 8 sources whose stellar parameters were calculated in this study, with essentially identical stellar parameters. After removing this source, stellar parameters from other 13 sources were incorporated. Since only 10 of these 13 sources had orbital radius/stellar mass data (with two sources having exclusively either orbital radius or stellar mass), the actual number of sources participating in calculations and plotting in this section was 10.

Through combined analysis of 21 RS CVn systems (8 from our parameter calculations plus 13 from \citet{toet2021}),  radio luminosity is found to exhibit significant positive correlations with primary star mass ($M_{\rm pri}$, PCC=0.663), total binary mass ($M_{\rm tot}$, PCC=0.663), and orbital semi-major axis ($a$, PCC=0.719). Meanwhile, no correlation is  found with secondary star mass or stellar activity quantified by Rossby number ($R_{\rm o}$).

The orbital semi-major axis exhibits the strongest correlation (PCC=0.719) with radio luminosity. This suggests that large-scale magnetic structures spanning the binary separation rather than individual stellar properties  are the primary factor governing radio emission in RS CVn systems. 

Further verification of this phenomenon requires more work. This can be multi-frequency monitoring such as long-term radio observations across MHz–GHz bands to precisely determine maximum luminosities. Alternatively, the parameter space can be expanded by cross-matching existing RS CVn catalogs with TESS data. If confirmed, the unified period-luminosity relationship challenges current models attributing radio emission solely to established mechanisms.

\begin{acknowledgments}

We are grateful to Drs. Min Dai, Zehao Zhang, Mengyao Xue, and Prof. Tara Murphy for their helpful discussion and suggestion. This work is supported by the NSFC project 12133002, and CMS-CSST-2021-A09.
\end{acknowledgments}

\bibliographystyle{aasjournal}
\bibliography{main.bib}

\begin{table}[]
    \centering
    \caption{The RS CVn radio data}
    \begin{tabular}{c c c c c c c c c c c}
    \hline \hline
        Name & SRSC & $P_{\rm rot}$ & $errP_{\rm rot}$ & $S_{\rm intI}$ & $errS_{\rm intI}$ & $S_{\rm peakI}$ & $errS_{\rm peakI}$ & RA & Dec & Dist \\

         & & day & day & mJy & mJy & mJy/beam & mJy/beam & $\degree$ & $\degree$ & pc \\
    \hline

V1198 Ori & 28 & 0.310 & 0.002 & - & - & 1.20 & 0.300 & 74.5701 & 0.4537 & 34 \\
V841 Cen & 30 & 5.980 & 0.341 & 12.00 & 2.000 & 7.60 & 0.700 & 218.5659 & -60.4082 & 95 \\
V376 Cep & 36 & 1.153 & 0.027 & - & - & 4.10 & 0.300 & 329.5773 & 82.8709 & 40 \\
EI Eri & 79 & 1.930 & 0.072 & 9.80 & 0.300 & 8.60 & 0.100 & 62.4206 & -7.8924 & 55 \\
HD 22468 & 81 & 2.840 & 0.160 & 80.30 & 0.100 & 78.17 & 0.100 & 54.1969 & 0.5870 & 30 \\
V835 Her & 86 & 3.320 & 0.115 & - & - & 0.50 & 0.100 & 268.8521 & 36.1888 & 31 \\
UX For & 190 & 0.950 & 0.017 & 2.30 & - & 1.90 & - & 40.8570 & -37.9288 & 41 \\
HR Psc & 221 & 3.940 & 0.323 & 4.05 & 0.030 & 4.14 & 0.010 & 24.1159 & 25.1431 & 218 \\
CD-56 1450 & 229 & 18.500 & 1.116 & 0.67 & 0.006 & 0.67 & 0.004 & 93.2749 & -56.3404 & 385 \\
UX Ari & 250 & 6.420 & 0.265 & 94.90 & 0.500 & 88.80 & 0.300 & 51.6476 & 28.7146 & 50 \\
BC Phe & 252 & 0.657 & 0.002 & 0.55 & 0.010 & 0.58 & 0.004 & 20.5790 & -56.7316 & 114 \\
HD 217344 & 255 & 1.648 & 0.054 & 3.99 & 0.060 & 3.92 & 0.020 & 345.1177 & -33.7457 & 83 \\
AE Men & 271 & 11.980 & 0.316 & 3.20 & 0.900 & 2.80 & 0.500 & 96.4178 & -72.0431 & 286 \\
HU Vir & 275 & 10.350 & 0.576 & 4.60 & 0.300 & 4.60 & 0.200 & 183.3362 & -9.0797 & 145 \\
HD 155555 & 281 & 1.680 & 0.053 & 7.60 & 0.400 & 7.20 & 0.200 & 259.3560 & -66.9516 & 30 \\
RX J0505.6-5755 & 299 & 0.632 & 0.003 & 0.23 & 0.007 & 0.18 & 0.001 & 76.4022 & -57.9266 & 94 \\
GK Hya & 313 & 3.588 & 0.064 & 0.43 & 0.010 & 0.40 & 0.004 & 127.7053 & 2.2741 & 232 \\
SZ Psc & 354 & 3.970 & 0.078 & 18.90 & 0.200 & 18.30 & 0.100 & 348.3492 & 2.6756 & 90 \\
AR Lac & 371 & 1.983 & 0.018 & 16.60 & 0.300 & 15.80 & 0.200 & 332.1697 & 45.7425 & 43 \\
V1265 Cen & 375 & 3.540 & 0.118 & 5.30 & 0.300 & 5.20 & 0.200 & 197.7298 & -48.7344 & 326 \\
CD-50 4879 & 380 & 4.530 & 0.207 & 1.60 & 0.300 & 1.40 & 0.100 & 152.7673 & -51.3300 & 221 \\
EV Dra & 385 & 1.660 & 0.028 & 3.70 & 0.300 & - & - & 240.4472 & 51.3479 & 57 \\
DM UMa & 404 & 7.300 & 0.938 & 4.10 & 0.300 & - & - & 163.9311 & 60.4693 & 188 \\
FF UMa & 423 & 3.280 & 0.049 & 1.90 & 0.300 & - & - & 143.4438 & 62.8278 & 114 \\
TY Pyx & 445 & 3.200 & 0.052 & 1.10 & 0.300 & 1.00 & 0.200 & 134.9278 & -27.8165 & 55 \\
ER Eri & 451 & 5.960 & 0.322 & 2.00 & 0.200 & 1.70 & 0.100 & 32.5333 & -54.5111 & 203 \\
SZ Pic & 490 & 4.950 & 0.058 & 1.30 & 0.300 & 1.40 & 0.200 & 88.3640 & -43.5589 & 187 \\
HD 150708A & 501 & 4.630 & 0.194 & 0.60 & 0.100 & - & - & 249.7668 & 60.6994 & 153 \\
BH CVn & 533 & 2.620 & 0.131 & 14.50 & - & 13.40 & - & 203.6997 & 37.1824 & 46 \\
V478 Lyr & 535 & 2.130 & 0.083 & 2.60 & 0.400 & 2.40 & 0.200 & 286.8855 & 30.2550 & 27 \\
V406 Gem & 595 & 11.200 & 1.343 & 1.40 & 0.300 & 1.20 & 0.200 & 101.8149 & 14.5775 & 688 \\
UX Com & 613 & 3.640 & 0.064 & 1.70 & 0.300 & 1.30 & 0.100 & 195.3873 & 28.6317 & 211 \\
VASC J1628-41 & 634 & 4.900 & 0.444 & 2.30 & 0.600 & 2.30 & 0.300 & 247.1970 & -41.8776 & 293 \\
EZ Eri & 656 & 8.800 & 1.650 & 1.10 & 0.400 & 1.10 & 0.200 & 76.8687 & -5.4072 & 264 \\
V1257 Cen & 667 & 2.760 & 0.149 & 1.00 & 0.300 & 1.00 & 0.200 & 192.7141 & -51.9432 & 109 \\
V498 And & 719 & 11.000 & 2.039 & 1.90 & 0.200 & 1.30 & 0.090 & 10.0870 & 43.7236 & 124 \\
CD-35 4072 & 725 & 2.370 & 0.057 & 1.50 & 0.300 & 1.30 & 0.200 & 119.7411 & -35.3715 & 238 \\
BD+14 2519 & 727 & 4.530 & 0.381 & 1.20 & 0.600 & 1.60 & 0.400 & 188.9891 & 13.4902 & 202 \\
VV Mon & 736 & 6.050 & 0.186 & 4.40 & 0.300 & 4.00 & 0.200 & 105.8262 & -5.7376 & 277 \\
IN Leo & 740 & 6.240 & 0.097 & 1.10 & 0.400 & 1.20 & 0.300 & 159.9959 & 13.4560 & 228 \\
SS Boo & 799 & 7.640 & 0.548 & 1.00 & 0.300 & 1.20 & 0.200 & 228.3853 & 38.5681 & 236 \\
BY Cet & 821 & 2.560 & 0.134 & 0.60 & 0.100 & 0.65 & 0.090 & 41.8623 & -0.2060 & 69 \\

\hline
    \end{tabular}
    \label{tab:RS Cvn radio data}
\end{table}

\begin{table}[]
\begin{threeparttable}
    \centering
    \caption{The RS CVn stellar parameter}
    \begin{tabular}{c c c c c c c c c c}
    \hline \hline
        Star name & $P_{\rm rot}$ & $a$ & $T_{\rm eff}^{\rm sec}/T_{\rm eff}^{\rm pri}$ & $T_{\rm eff}^{\rm pri}$ & $T_{\rm eff}^{\rm sec}$ & $M_*^{\rm pri}$ & $M_*^{\rm sec}$ & Luminosity & $R_{\rm o}$ \\

         & day & AU & & K & K & $\rm M_{\odot}$ & $\rm M_{\odot}$ & $\rm L_{\rm \odot}$ & \\
    \hline

        EV Dra & 1.660 & 0.034 & 0.930 & 5592 & 5201 & 0.961 & 0.872 & 1.22 & 0.094 \\
        BY Cet & 2.560 & 0.043 & 0.966 & 5095 & 4922 & 0.850 & 0.812 & 0.69 & 0.131 \\
        RX J0505.6-5755 & 0.632 & 0.016 & 0.949 & 4610 & 4375 & 0.750 & 0.643 & 0.31 & 0.024 \\
        UX For & 0.950 & 0.023 & 0.985 & 5448 & 5366 & 0.927 & 0.910 & 1.20 & 0.057 \\
        V1257 Cen & 2.760 & 0.044 & 0.968 & 4708 & 4558 & 0.770 & 0.739 & 0.39 & 0.124 \\
        V478 Lyr & 2.130 & 0.036 & 0.696 & 5303 & 3691 & 0.894 & 0.479 & 0.56 & 0.056 \\
        V1198 Ori & 0.310 & 0.011 & 0.957 & 5638 & 5396 & 0.971 & 0.916 & 1.41 & 0.019 \\
        V376 Cep & 1.153 & 0.027 & 0.939 & 5690 & 5343 & 0.978 & 0.905 & 1.43 & 0.069 \\
        \hline \hline
        Sig CrB (TZ Crb) & 1.157 & 0.028 & - & - & - & 1.110 & 1.080 & - & 0.090 \\
        BQ CVn & 18.500 & - & - & - & - & - & - & - & - \\
        FG UMa & 21.300 & 0.192 & - & - & - & 1.500 & 0.580 & - & 0.702 \\
        BF Lyn & 3.804 & 0.055 & - & - & - & 0.710 & 0.710 & - & 0.162 \\
        DM UMa & 7.471 & - & - & - & - & 1.200 & - & - & 0.683 \\
        DG CVn (GJ3789) & 0.108 & - & - & - & - & - & - & - & - \\
        EZ Peg & 11.660 & 0.124 & - & - & - & - & - & - & - \\
        BH CVn (HR 5110) & 2.613 & 0.050 & - & - & - & 1.500 & 0.420 & - & 0.059 \\
        WW Dra & 4.630 & 0.076 & - & - & - & 1.360 & 1.340 & - & 0.505 \\
        YY Gem & 0.814 & 0.026 & - & - & - & 0.600 & 0.590 & - & 0.027 \\
        II Peg & 6.725 & 0.074 & - & - & - & 0.800 & 0.400 & - & 0.145 \\
        BD+334462 & 10.121 & 0.111 & - & - & - & 0.890 & 0.910 & - & 0.590 \\
        FG Cam (HD 61396) & 31.950 & 0.315 & - & - & - & 3.270 & 0.380 & - & 0.651 \\
    \hline
    \end{tabular}
    \label{tab:Stellar parameter}
      \begin{tablenotes}
    \small
    \item Notes. This table is divided into two sections. The upper section presents eight sources and their corresponding data, for which stellar parameters were calculated in this study. The lower section contains thirteen sources and their data, which are taken from \citet{toet2021}
  \end{tablenotes}
\end{threeparttable}
\end{table}



\end{CJK}
\end{document}